\documentclass[prl,twocolumn,superscriptaddress]{revtex4-1}

\usepackage{amssymb,amsmath,amsfonts,bm}
\usepackage{graphicx}
\usepackage{epstopdf}
\usepackage{times}
\usepackage{float}
\usepackage{lipsum}
\usepackage{color}
\usepackage{hyperref}

\hypersetup{colorlinks=true, linkcolor=blue, urlcolor=blue, citecolor=black}

\begin{document}

\title{Quantum Control via Stimulated Raman User-defined Passage}

\author{Jingjing Niu}
\thanks{J. J. Niu, B.-J. L., and Y. X. Zhou contributed equally to this work.}
\affiliation{Institute for Quantum Science and Engineering and Department of Physics, Southern University of Science and Technology, Shenzhen 518055, China}
\author{Bao-Jie Liu}
\thanks{J. J. Niu, B.-J. L., and Y. X. Zhou contributed equally to this work.}
\affiliation{Institute for Quantum Science and Engineering and Department of Physics, Southern University of Science and Technology, Shenzhen 518055, China}
\author{Yuxuan Zhou}
\thanks{J. J. Niu, B.-J. L., and Y. X. Zhou contributed equally to this work.}
\affiliation{Institute for Quantum Science and Engineering and Department of Physics, Southern University of Science and Technology, Shenzhen 518055, China}
\author{Tongxing Yan}
\affiliation{Institute for Quantum Science and Engineering and Department of Physics, Southern University of Science and Technology, Shenzhen 518055, China}
\author{Wenhui Huang}
\affiliation{Institute for Quantum Science and Engineering and Department of Physics, Southern University of Science and Technology, Shenzhen 518055, China}
\author{Weiyang Liu}
\affiliation{Institute for Quantum Science and Engineering and Department of Physics, Southern University of Science and Technology, Shenzhen 518055, China}
\author{Libo Zhang}
\affiliation{Institute for Quantum Science and Engineering and Department of Physics, Southern University of Science and Technology, Shenzhen 518055, China}
\author{Hao Jia}
\affiliation{Institute for Quantum Science and Engineering and Department of Physics, Southern University of Science and Technology, Shenzhen 518055, China}
\author{Song Liu}
\email{lius3@sustech.edu.cn}
\affiliation{Institute for Quantum Science and Engineering and Department of Physics, Southern University of Science and Technology, Shenzhen 518055, China}
\author{Man-Hong Yung}
\email{yung@sustech.edu.cn}
\affiliation{Institute for Quantum Science and Engineering and Department of Physics, Southern University of Science and Technology, Shenzhen 518055, China}
\affiliation{Central Research Institute, Huawei Technologies, Shenzhen, 518129, China}
\author{Yuanzhen Chen}
\email{chenyz@sustech.edu.cn}
\affiliation{Institute for Quantum Science and Engineering and Department of Physics, Southern University of Science and Technology, Shenzhen 518055, China}
\author{Dapeng Yu}
\affiliation{Institute for Quantum Science and Engineering and Department of Physics, Southern University of Science and Technology, Shenzhen 518055, China}

\date{\today}

\begin{abstract}
Stimulated Raman adiabatic passage (STIRAP) is a widely-used technique of coherent state-to-state manipulation for many applications in physics, chemistry, and beyond. The adiabatic evolution of the state involved in STIRAP, called adiabatic passage, guarantees its robustness against control errors, but also leads to problems of low efficiency and decoherence.  Here we propose and experimentally demonstrate an alternative approach, termed stimulated Raman ``user-defined" passage (STIRUP), where a parameterized state is employed for constructing desired evolutions to replace the adiabatic passage in STIRAP. The user-defined passages can be flexibly designed for optimizing different objectives for different tasks, e.g. minimizing leakage error. To experimentally benchmark its performance, we apply STIRUP to the task of coherent state transfer in a superconducting Xmon qutrit. We found that STIRUP completed the transfer more then four times faster than STIRAP with enhanced robustness, and achieved a fidelity of 99.5\%, which is the highest among all recent experiments based on STIRAP and its variants. In practice, STIRUP differs from STIRAP only in the design of driving pulses; therefore, most existing applications of STIRAP can be readily implemented with STIRUP.
\end{abstract}

\maketitle

\emph{Introduction}.\textbf{---}Stimulated Raman adiabatic passage (STIRAP) was originally proposed for coherent population transfer between two uncoupled or weakly coupled quantum states via an intermediate state~\cite{Gaubatz1990,Kuklinski1989,Bergmann1998}. It is immune to loss through spontaneous emission of the intermediate state, and the adiabaticity guarantees its robustness against fluctuations in control parameters~\cite{Shore2008}. Due to such advantages, STIRAP and its many variants have quickly evolved into general methods of quantum manipulation beyond the original usage for state transfer, and found wide applications in many subfields of physics~\cite{Vitanov2017}, chemistry~\cite{Vitanov2001}, and engineering~\cite{Longhi2009}. One particular field is quantum information processing, where STIRAP has been proposed and demonstrated for implementing quantum gates~\cite{Duan2001,Kis2002,Unanyan2004,Pachos2004,Lacour2006,Moller2007,Menzel-Jones2007,Moller2008,Rousseaux2013}, state preparation and transfer~\cite{Unanyan2003,Linington2008,Premaratne2017,Kumar2016,Xu2016}, quantum computation in extended Hilbert space~\cite{Daems2007,Daems2008}, quantum memory~\cite{Klein2007,Alexander2008}, and so on. However, the demand of high precision and fidelity in this field, as well as in other fields where STIRAP may find applications~\cite{Vitanov2017}, poses great challenges to the original format of STIRAP. In particular, its requirement of adiabaticity leads to two issues. Firstly, the efficiency or fidelity can approach unity only in the long-time limit. Secondly, a slow passage is undesirable since decoherence is ubiquitous.

 To address such issues, variants of STIRAP have been proposed, such as pulse shaping~\cite{Vasilev2009,Dridi2009}, composite STIRAP~\cite{Torosov2013}, and shortcut-to-adiabaticity (STA)~\cite{Guery-Odelin2019,Unanyan1997,Demirplak2003,Demirplak2008,Berry2009,Chen2010,Du2016,An2016,Vepsalainen2019,Yang2019,Zhou2016,Baksic2016,Chen2012,Laforgue2019,Xu2019,Vepsalainen2019}. The former two methods aim at reducing nonadiabatic losses by using complex pulse engineering and thus sacrifice some of the major advantages of STIRAP: its simplicity and immunity to fluctuations in control parameters. STA represents a family of schemes aiming at accelerating adiabatic processes, including the original counterdiabatic driving~\cite{Berry2009,Chen2010,Du2016,An2016,Vepsalainen2019} and more sophisticated ones constructed via dressed states~\cite{Zhou2016,Baksic2016} or dynamical invariants~\cite{Chen2012,Laforgue2019,Xu2019}. The essential idea of counterdiabatic driving is to use an auxiliary drive to recast adiabatic paths. It requires a direct coupling between the initial and final states, which limits its application in many systems. On the other hand, methods based on dressed states~\cite{Baksic2016} or dynamical invariants~\cite{Chen2012} do not require direct coupling, but finding representations of dressed states or constructing proper dynamical invariants may become rather complicated, especially for high-dimensional quantum systems~\cite{Guery-Odelin2019}. These issues motivate us to search for an alternative way of improving STIRAP.

Here, we propose and experimentally demonstrate a new scheme of coherent quantum control, where a parameterized state is employed to construct desired passages. The scheme is termed stimulated Raman user-defined passage (STIRUP), which can be regarded as a generalization of STIRAP, as it yields identical result to STIRAP in the adiabatic limit. Furthermore, with the flexibility of defining passages, one can optimize different objectives for different tasks, such as minimizing leakage error, enhancing robustness against control errors, speeding up quantum control, etc.

For an experimental demonstration, the STIRUP pulses were applied to realize a coherent population transfer from the state $|0\rangle$ to $|2\rangle$ in an Xmon-type of superconducting qutrit. Previously, STIRAP was realized experimentally~\cite{Kumar2016,Xu2016,Premaratne2017} with a fidelity no more than 97\%, which is reproduced in our experiment. Recently, STA was also experimentally realized on a superconducting platform for the same task~\cite{Vepsalainen2019}, but only 96\% fidelity was achieved; mostly because of the additional microwave pulse for counterdiabatic driving. In contrast, both STIRAP and STIRUP require only two microwave pulses to complete the state transfer. Our experimental results show that STIRUP can be more than four times faster than STIRAP for the same fidelity, and the fidelity can reach $>$99.5\%, after pulse optimization against errors from leakage and cross coupling.

\begin{figure}[tbp]
	\centering\includegraphics[width=8.5cm]{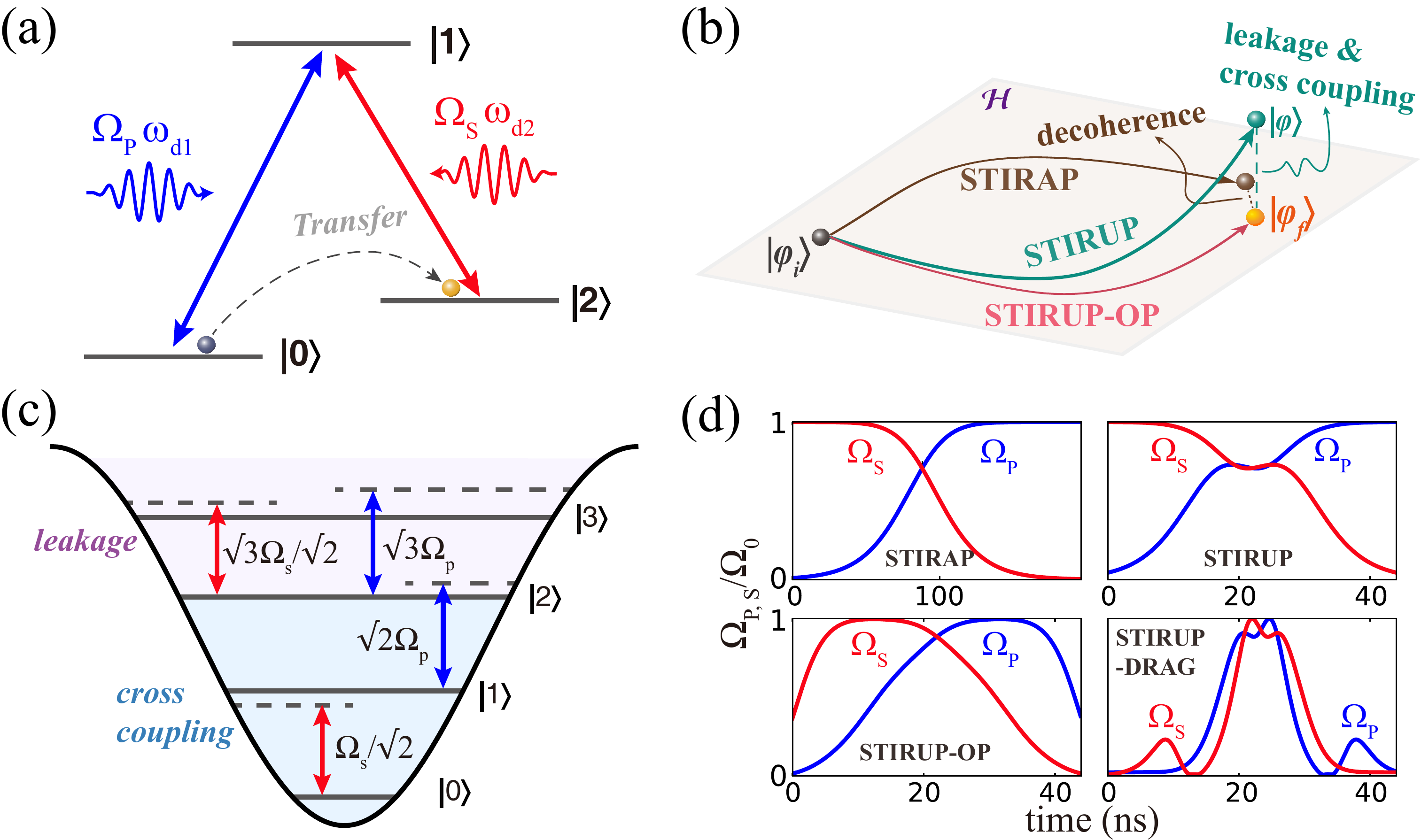}\caption{\label {setup} (a) Multilevel system driven by two external resonant pulses with time-dependent amplitudes. (b) Schematic of different passages. For multilevel systems with non-negligible cross coupling and leakage, unoptimized STIRUP may lead to significant leakage out of the relevant Hilbert space. (c) Cross coupling and leakage to higher excitations in a multilevel system with weak anharmonicity such as an Xmon type of superconducting device. For example, the Stokes pulse that resonantly couples $|1\rangle$ and $|2\rangle$ in (a) can now also introduce an off-resonant coupling between $|0\rangle$ and $|1\rangle$ in an Xmon. (d) Envelopes of microwave pulses used for the four cases studied in this work: STIRAP, STIRUP (no optimization), STIRUP-OP, and STIRUP-DRAG (both are optimized, see the main text for details).}\label{Fig1}
\end{figure}

\emph{Setting the stage}.\textbf{---}Let us consider a multi-level system under two external drives with time-dependent amplitudes, $\Omega_\mathrm{P}(t)$ and $\Omega_\mathrm{S}(t)$, as shown in Fig.\ref{setup}(a). Under the rotating-wave approximation, the Hamiltonian of this driven system reads $H_0(t) = \frac{1}{2}[\Omega_\mathrm{P}(t)|0\rangle\!\langle1| + \Omega_\mathrm{S}(t)|1\rangle\!\langle2| + \mathrm{H.c.}]$. One of its three eigenstates, $|D_0(t)\rangle \!=\!\cos\theta(t)|0\rangle + \sin\theta(t)|2\rangle$ $(\tan\theta(t)=\Omega_\mathrm{P}(t)/\Omega_\mathrm{S}(t))$, is a dark state, which is dynamically decoupled from the system evolution under the adiabatic condition~\cite{Vitanov2017}. STIRAP uses this dark state to realize a coherent population transfer from state $|0\rangle$ to $|2\rangle$ by evolving the mixing angle $\theta(t)$, without populating state $|1\rangle$~\cite{Vitanov2017}. In order to keep the system in the dark state, $\theta(t)$ is varied adiabatically to avoid transitions to other eigenstates of $H_0(t)$, which is called ``adiabatic passage".

In the following we maintain the setting of STIRAP, but provide a broader perspective on the use of ``passage" for quantum control. Specifically, here the concept of passage is extended by defining some time-parameterized state (as input) for inverse-engineering the driving Hamiltonian (as output). In this way, one gains higher flexibility in incorporating various optimizations for boosting performance. This strategy can also be applied to other quantum control applications beyond quantum state transfer.

For this purpose, our ``user-defined" passage for a three-level system can be generally parameterized as (up to a global phase):
\begin{equation}\label{PSI}
|\Phi_\text{UD}\rangle\!=\! \cos\!\gamma \cos\!\beta |0\rangle+e^{i \phi_{1}}\sin\!\gamma |1\rangle-e^{i \phi_{2}}\cos\!{\gamma}\sin\!\beta|2\rangle ,
\end{equation}
where $\beta(t)$, $\gamma(t)$, $\phi_{1,2}(t)$ are generally time-dependent variables to be determined below. To achieve state transfer from $|0\rangle$ to $|2\rangle$, it is sufficient to impose the following boundary conditions: $\gamma(0)=\gamma(T)=0$, $\beta(0)=0$, and $\beta(T)=\pi/2$ at time $t=0$ and $t=T>0$ respectively. Note that the STIRUP passage can be reduced to the adiabatic passage, and STIRUP becomes the same as STIRAP under the adiabatic condition.

Consequently, the time dependence of the control pulses can be determined by the passage  through the Schr\"{o}dinger equation,
\begin{equation}\label{IIRomega}
\begin{split}
\Omega_\mathrm{P}&= [\dot{\beta}\cot{\gamma}\sin\beta+\dot{\gamma}\cos\beta]e^{-i\phi}, \\
\Omega_\mathrm{S} &=[\dot{\beta}\cot\gamma\cos\beta-(\dot{\gamma}-i\dot{\phi}_{2}\cot{\gamma})\sin\beta]e^{i(\phi_{2}-\phi)},
\end{split}
\end{equation}
where $\phi\equiv\phi_{1}+\pi/2$. To eliminate the phase factors, we can choose $\phi=0$ and $\phi_{2}=0$ for simplicity. Note however that extra care must be taken at $t=0$ and $t=T$, as the boundary conditions would imply a divergence of the driving pulses whenever $\displaystyle \cot \gamma|_{\gamma \rightarrow 0}\rightarrow \infty $.

To overcome such a problem, it is sufficient to maintain the combination $G(t)\equiv \dot{\beta}(t)\cot{\gamma}(t)$ to be finite. In other words, on the user-defined passage we enforce additional boundary conditions: $G(0)\neq0$, $G(T)\neq0$, and $\dot{\beta}(0)=\dot{\beta}(T)=0$. As a result, we re-write the above relations as follows,
\begin{equation}\label{IRomega}
\begin{split}
\Omega_\mathrm{P}&=\sqrt{G^{2}+\dot{\gamma}^{2}} \sin [\beta+\arctan{(\dot{\gamma}/G)}], \\ \Omega_\mathrm{S}& =\sqrt{G^{2}+\dot{\gamma}^{2}} \cos [\beta+\arctan{(\dot{\gamma}/G)}] .
\end{split}
\end{equation}
To demonstrate the flexibility, we provide three possibles (i) choose $G(t)=\Omega_{0}$ to be some constant and $\beta(t)=\frac{\pi}{2}\frac{1}{1+e^{-10\tau/T}}$ with $\tau\equiv t-T/2$. (ii) To reduce the population of intermediate level $|1\rangle$, we can instead choose the parameter as $G(t)=\Omega_{0}[1+Ae^{-\frac{(\tau)^2}{(T/B)^2}}]$, where A and B can be determined via numerical optimization (see SM for details). (iii)  To minimize the cross coupling and leakage errors, one can also combine STIRUP with the hyper-Gaussian function $g(t)=\exp -(2 \tau)^{8}$ of the optimal STIRAP pulses~\cite{Vasilev2009}, i.e., $G(t)=\Omega_{0}g(t)[1+Ae^{-\frac{(\tau)^2}{(T/B)^2}}]$.
%Therefore, the realization of any specific manipulation of quantum states is translated into design of $\beta(t)$ and $G(t)$, namely, user-defined passages.

%We emphasize that, unlike STIRAP and many of its variants, the protocol introduced here does not impose any restriction (e.g., being adiabatic) on the temporal dynamics of the underlying evolution. Rather, it relies on directly engineering parameterized solutions of the Schr\"{o}dinger equation. Therefore, it is a general protocol that can reproduce the outcomes of all other methods by designing different passages (see SM Section I and II).

%In addition, if the adiabatic condition is imposed,  the STIRUP pulses can be reduced to the case of STIRAP. It can be shown that $|\Phi(t)\rangle$ becomes an instantaneous eigenstate of $H_0(t)$ of zero eigenenergy (dark state), and Eq. (\ref{IRomega}) reduces to  ${\Omega_\mathrm{P}=\Omega_{0}\sin \beta(t)}$ and ${\Omega_\mathrm{S}=\Omega_{0}\cos \beta(t)}$.

In the following, we discuss an exemplary implementation of STIRUP using superconducting quantum circuits (see Fig.\ref{Fig1}(c)). Specifically, we realized a state transfer process from $|0\rangle$ to $|2\rangle$ in an Xmon-type of superconducting qutrit. To the first order of approximation, this specific system can be viewed as an anharmonic oscillator, with an anharmonicity about one order of magnitude smaller than its characteristic frequency. Such a small anharmonicity renders the Hamiltonian $H_{0}(t)$ introduced above insufficient to describe an Xmon qutrit driven by two microwave pulses. Additional terms describing cross coupling and leakage to higher excited energy levels must be added to $H_{0}(t)$:
%\begin{equation}\small\label{Leakage}
%\begin{split}
%\!H(t)\! = &H_{0}(t)\!+\! \frac{1}{2}\!\!\left[\!\left(\!\frac{\Omega_\mathrm{S}}{\sqrt{2}}e\!^{\!-\!i\alpha t}|0\rangle\!\langle1| \!+\! \sqrt{2}\Omega_\mathrm{P} e\!^{\!-\!i\alpha t}|1\rangle\!\langle2|\!\right) \!+\! \mathrm{H.c.}\!\right]\\
%&\!+\! \frac{1}{2}\!\left[\!\left(\!\frac{\sqrt{6}\Omega_\mathrm{S}}{2}e\!^{\!-\!i\alpha t}\!+\!\sqrt{3}\Omega_\mathrm{P}e\!^{\!-\!i2\alpha t}\!\right)\!|2\rangle\!\langle3|\!+\! \mathrm{H.c.}\!\right]\ .
%\end{split}
%\end{equation}
\begin{eqnarray}\small\label{Leakage}
\!H(t)\! = &H_{0}(t)\!+\! \frac{1}{2}\!\!\left[\!\left(\!\frac{\Omega_\mathrm{S}}{\sqrt{2}}e\!^{\!-\!i\alpha t}|0\rangle\!\langle1| \!+\! \sqrt{2}\Omega_\mathrm{P} e\!^{\!-\!i\alpha t}|1\rangle\!\langle2|\!\right) \!+\! \mathrm{H.c.}\!\right] \nonumber\\
&\!+\! \frac{1}{2}\!\left[\!\left(\!\frac{\sqrt{6}\Omega_\mathrm{S}}{2}e\!^{\!-\!i\alpha t}\!+\!\sqrt{3}\Omega_\mathrm{P}e\!^{\!-\!i2\alpha t}\!\right)\!|2\rangle\!\langle3|\!+\! \mathrm{H.c.}\!\right]\ .
\end{eqnarray}

\begin{figure}[tbp]
	\centering
	\includegraphics[width=8.5cm]{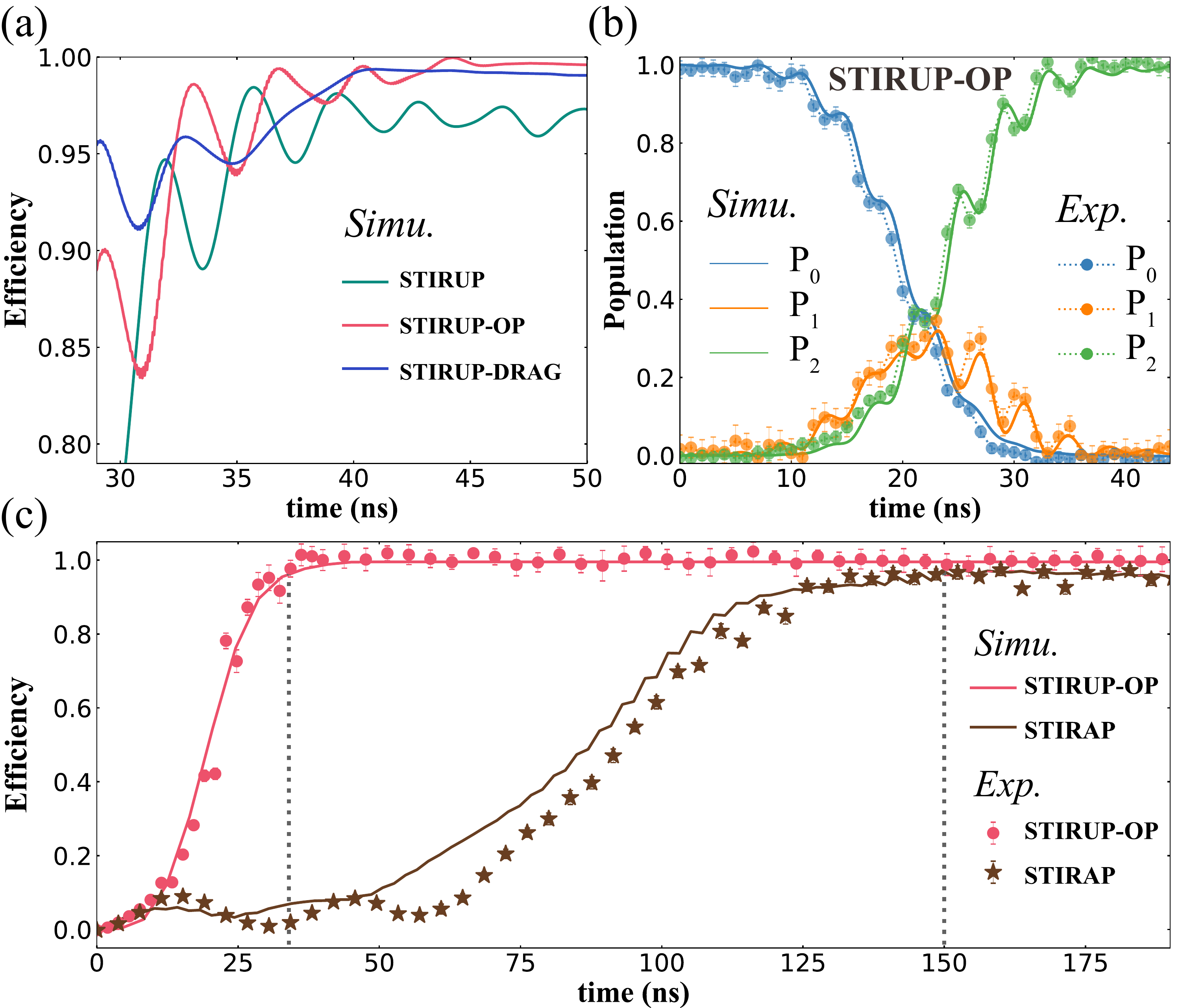}
	\caption{(a) Numerical simulation of the transfer efficiency (defined as the population of $|2\rangle$) towards the end of three different passages. The oscillatory behavior is due to the cross coupling and leakage discussed in the main text. (b) Evolution of populations during a state transfer using the STIRUP-OP passage. (c) Transfer efficiencies of two passages: STIRAP and STIRUP-OP. The two dotted lines at 34 and 150 ns mark the moments when an efficiency of 96\% is achieved for the STIRUP-OP and STIRAP passages, respectively.}\label{Fig2}
\end{figure}

Here we only consider the leakage to state $|3\rangle$, and $\alpha$ is the anharmonicity of the Xmon defined as the difference between the lowest two frequencies of transition: $\alpha=f_{12}-f_{01}$. With such additional contribution, the pulses designed based on Eq. (\ref{IRomega}) (obtained for $H_{0}(t)$) no longer give proper passages for the desired transfer. Indeed, numerical simulations indicate that the transfer process using such pulses exhibits a strong oscillatory behavior towards the end, which leads to ambiguity in evaluating the transfer efficiency (see the ``STIRUP" curve in Fig.\ref{Fig2}(a)). Such oscillatory behavior is due to the cross coupling and leakage discussed above. In the following we demonstrate that by modifying the pulses designed based on Eq. (\ref{IRomega}) using certain optimizations, one can effectively suppress these unwanted effects and achieve the desired state transfer with high efficiency and fidelity.

Two different ways of optimization are investigated. In the first one, we adapt the method used in Ref.~\cite{Davis1976,Vasilev2009} for optimizing the STIRAP process. In the second one, we extend the standard method of derivative removal by adiabatic gate (DRAG) for a three-level system to include even higher excited states~\cite{Motzoi2009,Gambetta2011,Chen2016}. In both optimizations, we run numerical simulations, using a master equation, to optimize the overall fidelity of the transfer process. Both optimized and unoptimized pulses (hereafter referred to as STIRUP-OP, STIRUP-DRAG, and STIRUP, respectively), together with those used in the STIRAP process, are compared in Fig. \ref{Fig1}(d). Further details of pulse design and optimization can be found in SM. With such optimizations, the oscillatory behavior mentioned above can be suppressed to a large extent. The transfer efficiency can now be determined to be 99.8\% and 99.5\% for the two optimizations, respectively (Fig.\ref{Fig2}(a)).

\emph{Experimental results and analysis}.\textbf{--} All data reported in this work were acquired on one Xmon qutrit, and similar results were reproduced on other samples. The characteristic frequencies of the qutrit used here are around $f_{10}=\omega_{10}/2\pi = 5.208$ GHz and $f_{21}=\omega_{21}/2\pi = 4.958$ GHz. The relaxation and dephasing times are $T_1^{10}=4.82$ $\mu$s, $T_2^{10}=5.06$ $\mu$s, $T_1^{21}=5.96$ $\mu$s and $T_2^{21}=2.55$ $\mu$s, respectively. External microwave drives are applied to the qutrit through an XY control line. The qutrit is capacitively coupled to a resonator of quarter-wavelength ($\omega_{r} /2\pi= 6.68~ \mathrm{GHz}$), which is in turn coupled to a transmission line. The state of the qutrit can be deduced by measuring the transmission coefficient $S_{21}$ of the transmission line using the dispersive readout scheme. Further details of the sample and measurement setup can be found in SM.

To characterize a state transfer process following a specific passage, we measure population of the three levels of the qutrit as a function of time. Figure \ref{Fig2}(b) shows a typical data set for the case of the STIRUP-OP passage. The qutrit is initialized to state $|0\rangle$, and the pulse shown on the third panel of Fig. \ref{Fig1}(d) is applied. Within 44 ns, the population of state $|2\rangle$ rises to above 99\%, accomplishing a fast and high fidelity transfer from state $|0\rangle$ to $|2\rangle$. Figure \ref{Fig2}(c) compares the efficiency of transfer (i.e., population of $|2\rangle$) for two different passages. For the STIRAP case, the maximum transfer efficiency (about 96\%) is achieved at around 150 ns, whereas the STIRUP-OP passage can reach the same efficiency within 34 ns. Such acceleration helps minimize losses due to spontaneous emission from excited states. As a result, the STIRUP can still achieve high fidelity even though the population of the intermediate state is nonzero during the transfer, unlike in STIRAP there is no occupation of the intermediate state during the whole process. We also emphasize that while the STIRUP passages used here do not specifically target at accelerating the transfer, like those STA-based variants of STIRAP, they nevertheless accomplish this goal nicely.

\begin{figure}[htbp]
	\includegraphics[width=8.5cm]{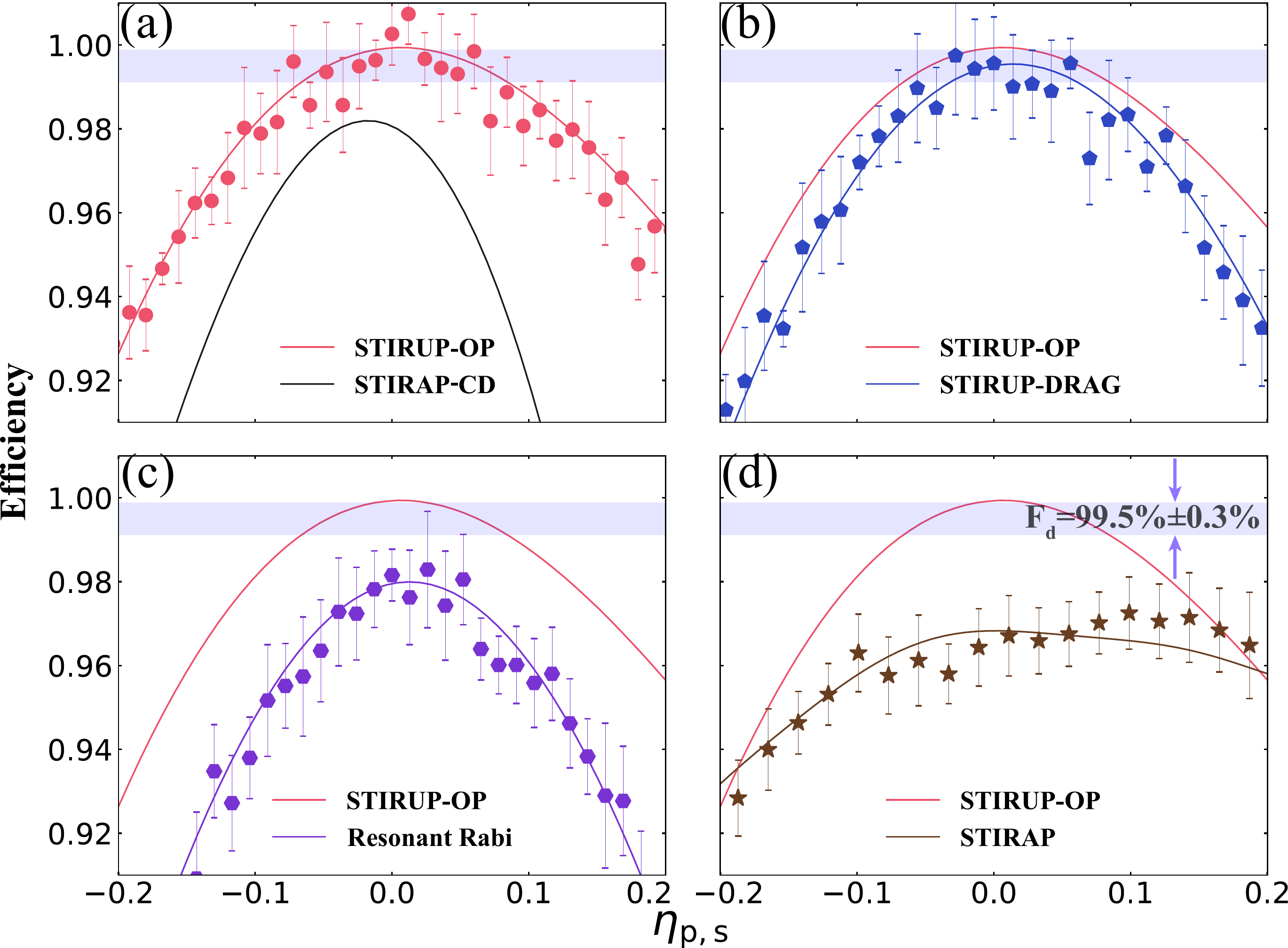}
	\caption{Robustness of the transfer efficiency against error in the Rabi frequency for five different passages. In all four panels, lines are results of numerical simulations, symbols are experimental results, and the shaded stripes indicate a range of the efficiency of (99.5 $\pm$ 0.3)\%. For comparison, the results of STIRUP-OP passage are shown in all four panels. In panel (a), the black line represents numerical simulation of a state transfer using STIRAP accelerated by a counterdiabatic driving, as reported in Ref~\cite{Vepsalainen2019}.}\label{Fig3}
\end{figure}

Next, We investigate the robustness of different protocols against experimental errors. In Fig. \ref{Fig3}, the transfer efficiencies of five passages against the error in the Rabi frequency $\eta_\mathrm{p,s}$ are compared to each other. $\eta_\mathrm{p,s}$ is defined as
$\tilde{\Omega}_\mathrm{P,S}=(1+\eta_\mathrm{p,s})\Omega_\mathrm{P,S}$, where $\tilde{\Omega}_\mathrm{P,S}$ and $\Omega_\mathrm{P,S}$ are actual and expected values of the Rabi frequency, respectively. For the resonant Rabi (RR) method~\cite{Christandl2004,Li2018a}, two identical $\pi$ pulses in the form of $\Omega_\mathrm{P,S}=\Omega_0\sin{\beta(t)}$ are applied. The RR scheme is often used to benchmark protocols of state transfer and certain quantum gates. The STIRUP-OP process has the best robustness in transfer efficiency against the error, larger than 92\% in the range of $0.8\!\sim\!1.2\Omega_\mathrm{P,S}$. Overall, the STIRUP-OP passage is superior to the RR scheme, the STIRAP cases (both original and accelerated forms), and the STIRUP-DRAG passage, in terms of both robustness and optimal transfer efficiency.

\begin{figure}[htbp]
	\includegraphics[width=8.5cm]{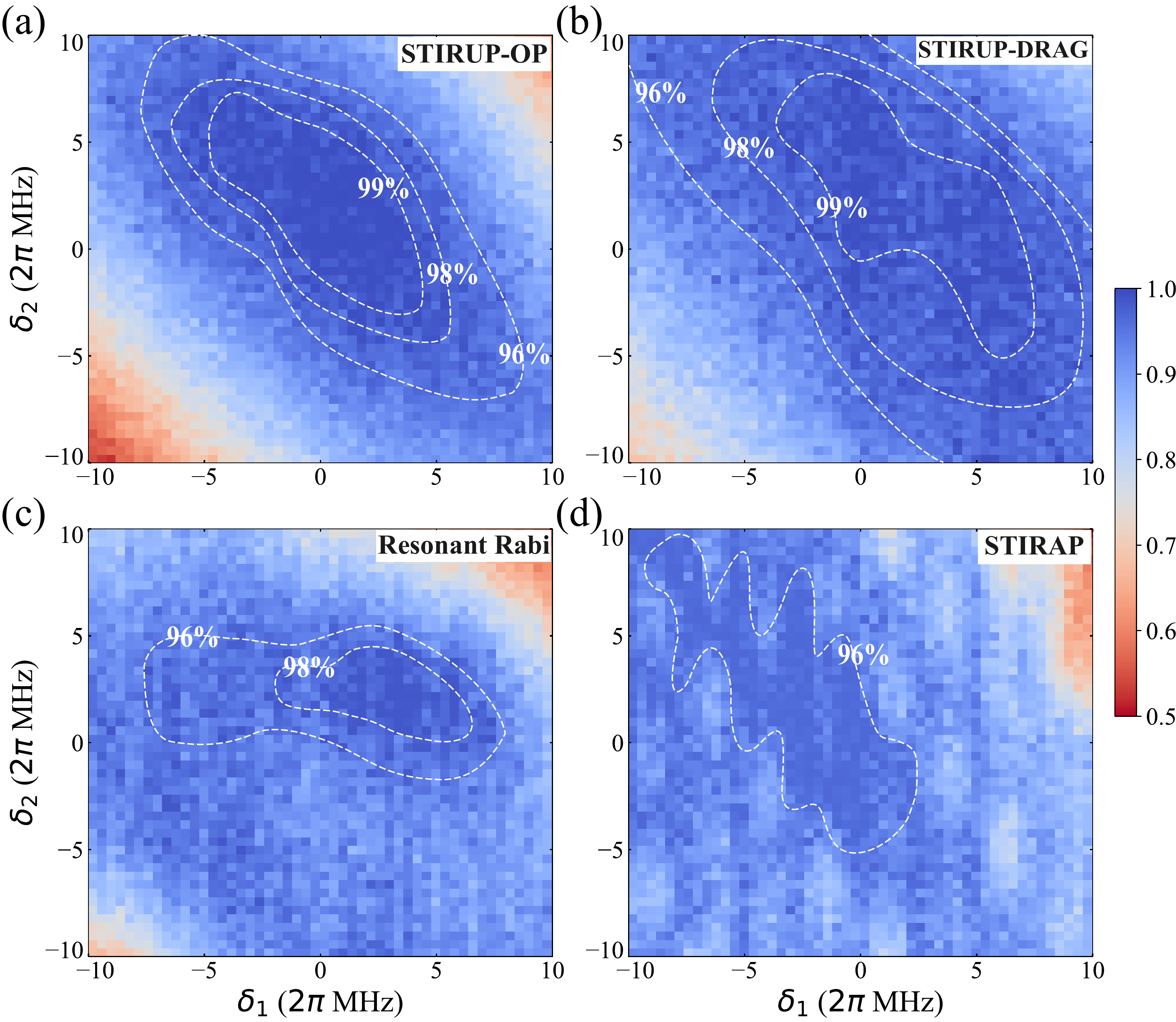}
	\caption{Robustness of the transfer efficiency against detuning errors $\delta_{1,2}$ for four different passages: (a) STIRUP-OP, (b) STIRUP-DRAG, (c) Resonant Rabi and (d) STIRAP. Dashed lines are contours indicating different values of the efficiency. All four panels are experimental results, and their comparison to numerical simulations can be found in SM.}\label{Fig4}
\end{figure}

We have also investigated, both numerically and experimentally, the effect of detuning in the pulse frequencies on the transfer efficiency of different methods. Figure \ref{Fig4} plots the transfer efficiency as a function of the two detunings defined by $\delta_1=\omega_\mathrm{d1}-\omega_{10}$ and $\delta_2=\omega_\mathrm{d2}-\omega_{21}$ ($\omega_\mathrm{d1}$ and $\omega_\mathrm{d2}$ are frequencies of the two external drives, see Fig. \ref{Fig1}(a)). The two methods using our optimization show similar performance. In both cases, the transfer efficiency is quite robust against the two detunings as long as their sum is kept near zero (i.e., the two-photon resonant condition is approximately held). Contours of the transfer efficiency can be easily identified on the experimental data, and are in reasonable agreement with the numerical simulations (see SM for a comparison). Overall, the RR and the adiabatic passage cases exhibit less robust performance against the detuning errors. Moreover, we have also run numerical simulations to compare our protocols to the scheme of STIRAP accelerated by a counterdiabatic driving~\cite{Vepsalainen2019}, and found a better robustness in the transfer efficiency against experimental errors for our protocols (see SM).

In summary, we have proposed and demonstrated STIRUP as a new protocol for quantum control. In this protocol, user-defined passages for realizing any specified quantum control are constructed via direct engineering of parameterized solutions of the Schr\"{o}dinger equation. As a demonstration of the protocol, we realized a coherent population transfer from the state $|0\rangle$ to $|2\rangle$ in an Xmon superconducting qutrit. Our scheme uses only two simple pulses, and can achieve a high fidelity ($>$99.5\%) in an efficient way, with excellent robustness against experimental errors. Performance wise, STIRUP is superior to existing methods in precision, efficiency, and robustness. In terms of simplicity, designing STIRUP pulses is rather straightforward, avoiding the complexity in various STA-based schemes. Such advantage of STIRUP becomes even more prominent for quantum control in systems of higher dimensions~\cite{Liu2019a}.

Although in this work a state transfer process in a three-level system is used to demonstrate the protocol, it can be generalized for more sophisticated applications. For example, some of the authors of this work have shown theoretically that STIRUP can be used for state transfer and one-step generation of entanglement in multi-level systems~\cite{Liu2019a}. In both cases, STIRUP achieves a significant improvement, in terms of precision, efficiency, and robustness, over STIRAP and other conventional methods. Besides quantum information processing, STIRUP may also find applications in other fields, such as precision measurement~\cite{Panda2016}, where the original STIRAP fails to deliver the high fidelity demanded. In general, we expect that STIRUP can replace STIRAP for most applications, while deliver enhanced performance in most cases.

The essential idea of STIRUP does not depend on assumptions specific to particular systems, so it should be readily implementable on other platforms. In the current work, STIRUP needs to be optimized to eliminate the cross coupling and leakage errors prevailing in Xmon type of superconducting devices, due to their relatively small anharmonicity. It therefore should be expected that on other platforms with larger anharmonicity, the application of STIRUP could be even more straightforward. Meanwhile, there always exists the flexibility of incorporating optimizations designed for specific systems. Finally, we note that the idea of using parametrized states for solving Schr\"{o}dinger equation can be extended to non-linear systems; for some cases, the population of the intermediate state must be non-zero~\cite{Dorier2017}.

\begin{acknowledgements}
We thank S. Gu\'{e}rin for a valuable discussion on our manuscript. This work was supported by the Key-Area Research and Development Program of GuangDong Province (Grant No. 2018B030326001), the National Natural Science Foundation of China (U1801661), the Guangdong Innovative and Entrepreneurial Research Team Program (2016ZT06D348), the Natural Science Foundation of Guangdong Province (2017B030308003), and the Science, Technology and Innovation Commission of Shenzhen Municipality (ZDSYS20170303165926217, JCYJ20170412152620376, KYTDPT20181011104202253).
\end{acknowledgements}


\begin{thebibliography}{52}%
\makeatletter
\providecommand \@ifxundefined [1]{%
 \@ifx{#1\undefined}
}%
\providecommand \@ifnum [1]{%
 \ifnum #1\expandafter \@firstoftwo
 \else \expandafter \@secondoftwo
 \fi
}%
\providecommand \@ifx [1]{%
 \ifx #1\expandafter \@firstoftwo
 \else \expandafter \@secondoftwo
 \fi
}%
\providecommand \natexlab [1]{#1}%
\providecommand \enquote  [1]{``#1''}%
\providecommand \bibnamefont  [1]{#1}%
\providecommand \bibfnamefont [1]{#1}%
\providecommand \citenamefont [1]{#1}%
\providecommand \href@noop [0]{\@secondoftwo}%
\providecommand \href [0]{\begingroup \@sanitize@url \@href}%
\providecommand \@href[1]{\@@startlink{#1}\@@href}%
\providecommand \@@href[1]{\endgroup#1\@@endlink}%
\providecommand \@sanitize@url [0]{\catcode `\\12\catcode `\$12\catcode
  `\&12\catcode `\#12\catcode `\^12\catcode `\_12\catcode `\%12\relax}%
\providecommand \@@startlink[1]{}%
\providecommand \@@endlink[0]{}%
\providecommand \url  [0]{\begingroup\@sanitize@url \@url }%
\providecommand \@url [1]{\endgroup\@href {#1}{\urlprefix }}%
\providecommand \urlprefix  [0]{URL }%
\providecommand \Eprint [0]{\href }%
\providecommand \doibase [0]{http://dx.doi.org/}%
\providecommand \selectlanguage [0]{\@gobble}%
\providecommand \bibinfo  [0]{\@secondoftwo}%
\providecommand \bibfield  [0]{\@secondoftwo}%
\providecommand \translation [1]{[#1]}%
\providecommand \BibitemOpen [0]{}%
\providecommand \bibitemStop [0]{}%
\providecommand \bibitemNoStop [0]{.\EOS\space}%
\providecommand \EOS [0]{\spacefactor3000\relax}%
\providecommand \BibitemShut  [1]{\csname bibitem#1\endcsname}%
\let\auto@bib@innerbib\@empty
%</preamble>
\bibitem [{\citenamefont {Gaubatz}\ \emph {et~al.}(1990)\citenamefont
  {Gaubatz}, \citenamefont {Rudecki}, \citenamefont {Schiemann},\ and\
  \citenamefont {Bergmann}}]{Gaubatz1990}%
  \BibitemOpen
  \bibfield  {author} {\bibinfo {author} {\bibfnamefont {U.}~\bibnamefont
  {Gaubatz}}, \bibinfo {author} {\bibfnamefont {P.}~\bibnamefont {Rudecki}},
  \bibinfo {author} {\bibfnamefont {S.}~\bibnamefont {Schiemann}}, \ and\
  \bibinfo {author} {\bibfnamefont {K.}~\bibnamefont {Bergmann}},\ }\href
  {\doibase 10.1063/1.458514} {\bibfield  {journal} {\bibinfo  {journal} {J.
  Chem. Phys.}\ }\textbf {\bibinfo {volume} {92}},\ \bibinfo {pages} {5363}
  (\bibinfo {year} {1990})}\BibitemShut {NoStop}%
\bibitem [{\citenamefont {Kuklinski}\ \emph {et~al.}(1989)\citenamefont
  {Kuklinski}, \citenamefont {Gaubatz}, \citenamefont {Hioe},\ and\
  \citenamefont {Bergmann}}]{Kuklinski1989}%
  \BibitemOpen
  \bibfield  {author} {\bibinfo {author} {\bibfnamefont {J.~R.}\ \bibnamefont
  {Kuklinski}}, \bibinfo {author} {\bibfnamefont {U.}~\bibnamefont {Gaubatz}},
  \bibinfo {author} {\bibfnamefont {F.~T.}\ \bibnamefont {Hioe}}, \ and\
  \bibinfo {author} {\bibfnamefont {K.}~\bibnamefont {Bergmann}},\ }\href
  {\doibase 10.1103/PhysRevA.40.6741} {\bibfield  {journal} {\bibinfo
  {journal} {Phys. Rev. A}\ }\textbf {\bibinfo {volume} {40}},\ \bibinfo
  {pages} {6741} (\bibinfo {year} {1989})}\BibitemShut {NoStop}%
\bibitem [{\citenamefont {Bergmann}\ \emph {et~al.}(1998)\citenamefont
  {Bergmann}, \citenamefont {Theuer},\ and\ \citenamefont
  {Shore}}]{Bergmann1998}%
  \BibitemOpen
  \bibfield  {author} {\bibinfo {author} {\bibfnamefont {K.}~\bibnamefont
  {Bergmann}}, \bibinfo {author} {\bibfnamefont {H.}~\bibnamefont {Theuer}}, \
  and\ \bibinfo {author} {\bibfnamefont {B.~W.}\ \bibnamefont {Shore}},\ }\href
  {\doibase 10.1103/RevModPhys.70.1003} {\bibfield  {journal} {\bibinfo
  {journal} {Rev. Mod. Phys.}\ }\textbf {\bibinfo {volume} {70}},\ \bibinfo
  {pages} {1003} (\bibinfo {year} {1998})}\BibitemShut {NoStop}%
\bibitem [{\citenamefont {Shore}(2008)}]{Shore2008}%
  \BibitemOpen
  \bibfield  {author} {\bibinfo {author} {\bibfnamefont {B.}~\bibnamefont
  {Shore}},\ }\href {\doibase 10.2478/v10155-010-0090-z} {\bibfield  {journal}
  {\bibinfo  {journal} {Acta Phys. Slovaca}\ }\textbf {\bibinfo {volume} {58}}
  (\bibinfo {year} {2008}),\ 10.2478/v10155-010-0090-z}\BibitemShut {NoStop}%
\bibitem [{\citenamefont {Vitanov}\ \emph {et~al.}(2017)\citenamefont
  {Vitanov}, \citenamefont {Rangelov}, \citenamefont {Shore},\ and\
  \citenamefont {Bergmann}}]{Vitanov2017}%
  \BibitemOpen
  \bibfield  {author} {\bibinfo {author} {\bibfnamefont {N.~V.}\ \bibnamefont
  {Vitanov}}, \bibinfo {author} {\bibfnamefont {A.~A.}\ \bibnamefont
  {Rangelov}}, \bibinfo {author} {\bibfnamefont {B.~W.}\ \bibnamefont {Shore}},
  \ and\ \bibinfo {author} {\bibfnamefont {K.}~\bibnamefont {Bergmann}},\
  }\href {\doibase 10.1103/RevModPhys.89.015006} {\bibfield  {journal}
  {\bibinfo  {journal} {Rev. Mod. Phys.}\ }\textbf {\bibinfo {volume} {89}},\
  \bibinfo {pages} {015006} (\bibinfo {year} {2017})}\BibitemShut {NoStop}%
\bibitem [{\citenamefont {Vitanov}\ \emph {et~al.}(2001)\citenamefont
  {Vitanov}, \citenamefont {Halfmann}, \citenamefont {Shore},\ and\
  \citenamefont {Bergmann}}]{Vitanov2001}%
  \BibitemOpen
  \bibfield  {author} {\bibinfo {author} {\bibfnamefont {N.}~\bibnamefont
  {Vitanov}}, \bibinfo {author} {\bibfnamefont {T.}~\bibnamefont {Halfmann}},
  \bibinfo {author} {\bibfnamefont {B.}~\bibnamefont {Shore}}, \ and\ \bibinfo
  {author} {\bibfnamefont {K.}~\bibnamefont {Bergmann}},\ }\href {\doibase
  10.1146/annurev.physchem.52.1.763} {\bibfield  {journal} {\bibinfo  {journal}
  {Annu. Rev. Phys. Chem.}\ }\textbf {\bibinfo {volume} {52}},\ \bibinfo
  {pages} {763} (\bibinfo {year} {2001})}\BibitemShut {NoStop}%
\bibitem [{\citenamefont {Longhi}(2009)}]{Longhi2009}%
  \BibitemOpen
  \bibfield  {author} {\bibinfo {author} {\bibfnamefont {S.}~\bibnamefont
  {Longhi}},\ }\href {\doibase 10.1002/lpor.200810055} {\bibfield  {journal}
  {\bibinfo  {journal} {Laser \& Photon. Rev.}\ }\textbf {\bibinfo {volume}
  {3}},\ \bibinfo {pages} {243} (\bibinfo {year} {2009})}\BibitemShut {NoStop}%
\bibitem [{\citenamefont {Duan}\ \emph {et~al.}(2001)\citenamefont {Duan},
  \citenamefont {Cirac},\ and\ \citenamefont {Zoller}}]{Duan2001}%
  \BibitemOpen
  \bibfield  {author} {\bibinfo {author} {\bibfnamefont {L.-M.}\ \bibnamefont
  {Duan}}, \bibinfo {author} {\bibfnamefont {J.~I.}\ \bibnamefont {Cirac}}, \
  and\ \bibinfo {author} {\bibfnamefont {P.}~\bibnamefont {Zoller}},\ }\href
  {\doibase 10.1126/science.1058835} {\bibfield  {journal} {\bibinfo  {journal}
  {Science}\ }\textbf {\bibinfo {volume} {292}},\ \bibinfo {pages} {1695}
  (\bibinfo {year} {2001})}\BibitemShut {NoStop}%
\bibitem [{\citenamefont {Kis}\ and\ \citenamefont {Renzoni}(2002)}]{Kis2002}%
  \BibitemOpen
  \bibfield  {author} {\bibinfo {author} {\bibfnamefont {Z.}~\bibnamefont
  {Kis}}\ and\ \bibinfo {author} {\bibfnamefont {F.}~\bibnamefont {Renzoni}},\
  }\href {\doibase 10.1103/PhysRevA.65.032318} {\bibfield  {journal} {\bibinfo
  {journal} {Phys. Rev. A}\ }\textbf {\bibinfo {volume} {65}},\ \bibinfo
  {pages} {032318} (\bibinfo {year} {2002})}\BibitemShut {NoStop}%
\bibitem [{\citenamefont {Unanyan}\ and\ \citenamefont
  {Fleischhauer}(2004)}]{Unanyan2004}%
  \BibitemOpen
  \bibfield  {author} {\bibinfo {author} {\bibfnamefont {R.~G.}\ \bibnamefont
  {Unanyan}}\ and\ \bibinfo {author} {\bibfnamefont {M.}~\bibnamefont
  {Fleischhauer}},\ }\href {\doibase 10.1103/PhysRevA.69.050302} {\bibfield
  {journal} {\bibinfo  {journal} {Phys. Rev. A}\ }\textbf {\bibinfo {volume}
  {69}},\ \bibinfo {pages} {050302} (\bibinfo {year} {2004})}\BibitemShut
  {NoStop}%
\bibitem [{\citenamefont {Pachos}\ and\ \citenamefont
  {Beige}(2004)}]{Pachos2004}%
  \BibitemOpen
  \bibfield  {author} {\bibinfo {author} {\bibfnamefont {J.~K.}\ \bibnamefont
  {Pachos}}\ and\ \bibinfo {author} {\bibfnamefont {A.}~\bibnamefont {Beige}},\
  }\href {\doibase 10.1103/PhysRevA.69.033817} {\bibfield  {journal} {\bibinfo
  {journal} {Phys. Rev. A}\ }\textbf {\bibinfo {volume} {69}},\ \bibinfo
  {pages} {033817} (\bibinfo {year} {2004})}\BibitemShut {NoStop}%
\bibitem [{\citenamefont {Lacour}\ \emph {et~al.}(2006)\citenamefont {Lacour},
  \citenamefont {Sangouard}, \citenamefont {Gu\'erin},\ and\ \citenamefont
  {Jauslin}}]{Lacour2006}%
  \BibitemOpen
  \bibfield  {author} {\bibinfo {author} {\bibfnamefont {X.}~\bibnamefont
  {Lacour}}, \bibinfo {author} {\bibfnamefont {N.}~\bibnamefont {Sangouard}},
  \bibinfo {author} {\bibfnamefont {S.}~\bibnamefont {Gu\'erin}}, \ and\
  \bibinfo {author} {\bibfnamefont {H.~R.}\ \bibnamefont {Jauslin}},\ }\href
  {\doibase 10.1103/PhysRevA.73.042321} {\bibfield  {journal} {\bibinfo
  {journal} {Phys. Rev. A}\ }\textbf {\bibinfo {volume} {73}},\ \bibinfo
  {pages} {042321} (\bibinfo {year} {2006})}\BibitemShut {NoStop}%
\bibitem [{\citenamefont {M\o{}ller}\ \emph {et~al.}(2007)\citenamefont
  {M\o{}ller}, \citenamefont {S\o{}rensen}, \citenamefont {Thomsen},\ and\
  \citenamefont {Drewsen}}]{Moller2007}%
  \BibitemOpen
  \bibfield  {author} {\bibinfo {author} {\bibfnamefont {D.}~\bibnamefont
  {M\o{}ller}}, \bibinfo {author} {\bibfnamefont {J.~L.}\ \bibnamefont
  {S\o{}rensen}}, \bibinfo {author} {\bibfnamefont {J.~B.}\ \bibnamefont
  {Thomsen}}, \ and\ \bibinfo {author} {\bibfnamefont {M.}~\bibnamefont
  {Drewsen}},\ }\href {\doibase 10.1103/PhysRevA.76.062321} {\bibfield
  {journal} {\bibinfo  {journal} {Phys. Rev. A}\ }\textbf {\bibinfo {volume}
  {76}},\ \bibinfo {pages} {062321} (\bibinfo {year} {2007})}\BibitemShut
  {NoStop}%
\bibitem [{\citenamefont {Menzel-Jones}\ and\ \citenamefont
  {Shapiro}(2007)}]{Menzel-Jones2007}%
  \BibitemOpen
  \bibfield  {author} {\bibinfo {author} {\bibfnamefont {C.}~\bibnamefont
  {Menzel-Jones}}\ and\ \bibinfo {author} {\bibfnamefont {M.}~\bibnamefont
  {Shapiro}},\ }\href {\doibase 10.1103/PhysRevA.75.052308} {\bibfield
  {journal} {\bibinfo  {journal} {Phys. Rev. A}\ }\textbf {\bibinfo {volume}
  {75}},\ \bibinfo {pages} {052308} (\bibinfo {year} {2007})}\BibitemShut
  {NoStop}%
\bibitem [{\citenamefont {M\o{}ller}\ \emph {et~al.}(2008)\citenamefont
  {M\o{}ller}, \citenamefont {Madsen},\ and\ \citenamefont
  {M\o{}lmer}}]{Moller2008}%
  \BibitemOpen
  \bibfield  {author} {\bibinfo {author} {\bibfnamefont {D.}~\bibnamefont
  {M\o{}ller}}, \bibinfo {author} {\bibfnamefont {L.~B.}\ \bibnamefont
  {Madsen}}, \ and\ \bibinfo {author} {\bibfnamefont {K.}~\bibnamefont
  {M\o{}lmer}},\ }\href {\doibase 10.1103/PhysRevLett.100.170504} {\bibfield
  {journal} {\bibinfo  {journal} {Phys. Rev. Lett.}\ }\textbf {\bibinfo
  {volume} {100}},\ \bibinfo {pages} {170504} (\bibinfo {year}
  {2008})}\BibitemShut {NoStop}%
\bibitem [{\citenamefont {Rousseaux}\ \emph {et~al.}(2013)\citenamefont
  {Rousseaux}, \citenamefont {Gu\'erin},\ and\ \citenamefont
  {Vitanov}}]{Rousseaux2013}%
  \BibitemOpen
  \bibfield  {author} {\bibinfo {author} {\bibfnamefont {B.}~\bibnamefont
  {Rousseaux}}, \bibinfo {author} {\bibfnamefont {S.}~\bibnamefont {Gu\'erin}},
  \ and\ \bibinfo {author} {\bibfnamefont {N.~V.}\ \bibnamefont {Vitanov}},\
  }\href {\doibase 10.1103/PhysRevA.87.032328} {\bibfield  {journal} {\bibinfo
  {journal} {Phys. Rev. A}\ }\textbf {\bibinfo {volume} {87}},\ \bibinfo
  {pages} {032328} (\bibinfo {year} {2013})}\BibitemShut {NoStop}%
\bibitem [{\citenamefont {Unanyan}\ and\ \citenamefont
  {Fleischhauer}(2003)}]{Unanyan2003}%
  \BibitemOpen
  \bibfield  {author} {\bibinfo {author} {\bibfnamefont {R.~G.}\ \bibnamefont
  {Unanyan}}\ and\ \bibinfo {author} {\bibfnamefont {M.}~\bibnamefont
  {Fleischhauer}},\ }\href {\doibase 10.1103/PhysRevLett.90.133601} {\bibfield
  {journal} {\bibinfo  {journal} {Phys. Rev. Lett.}\ }\textbf {\bibinfo
  {volume} {90}},\ \bibinfo {pages} {133601} (\bibinfo {year}
  {2003})}\BibitemShut {NoStop}%
\bibitem [{\citenamefont {Linington}\ and\ \citenamefont
  {Vitanov}(2008)}]{Linington2008}%
  \BibitemOpen
  \bibfield  {author} {\bibinfo {author} {\bibfnamefont {I.~E.}\ \bibnamefont
  {Linington}}\ and\ \bibinfo {author} {\bibfnamefont {N.~V.}\ \bibnamefont
  {Vitanov}},\ }\href {\doibase 10.1103/PhysRevA.77.062327} {\bibfield
  {journal} {\bibinfo  {journal} {Phys. Rev. A}\ }\textbf {\bibinfo {volume}
  {77}},\ \bibinfo {pages} {062327} (\bibinfo {year} {2008})}\BibitemShut
  {NoStop}%
\bibitem [{\citenamefont {Premaratne}\ \emph {et~al.}(2017)\citenamefont
  {Premaratne}, \citenamefont {Wellstood},\ and\ \citenamefont
  {Palmer}}]{Premaratne2017}%
  \BibitemOpen
  \bibfield  {author} {\bibinfo {author} {\bibfnamefont {S.~P.}\ \bibnamefont
  {Premaratne}}, \bibinfo {author} {\bibfnamefont {F.~C.}\ \bibnamefont
  {Wellstood}}, \ and\ \bibinfo {author} {\bibfnamefont {B.~S.}\ \bibnamefont
  {Palmer}},\ }\href {https://doi.org/10.1038/ncomms14148} {\bibfield
  {journal} {\bibinfo  {journal} {Nat. Commun.}\ }\textbf {\bibinfo {volume}
  {8}},\ \bibinfo {pages} {14148} (\bibinfo {year} {2017})}\BibitemShut
  {NoStop}%
\bibitem [{\citenamefont {Kumar}\ \emph {et~al.}(2016)\citenamefont {Kumar},
  \citenamefont {Veps\"{a}l\"{a}inen}, \citenamefont {Danilin},\ and\
  \citenamefont {Paraoanu}}]{Kumar2016}%
  \BibitemOpen
  \bibfield  {author} {\bibinfo {author} {\bibfnamefont {K.~S.}\ \bibnamefont
  {Kumar}}, \bibinfo {author} {\bibfnamefont {A.}~\bibnamefont
  {Veps\"{a}l\"{a}inen}}, \bibinfo {author} {\bibfnamefont {S.}~\bibnamefont
  {Danilin}}, \ and\ \bibinfo {author} {\bibfnamefont {G.~S.}\ \bibnamefont
  {Paraoanu}},\ }\href {https://doi.org/10.1038/ncomms10628} {\bibfield
  {journal} {\bibinfo  {journal} {Nat. Commun.}\ }\textbf {\bibinfo {volume}
  {7}},\ \bibinfo {pages} {10628} (\bibinfo {year} {2016})}\BibitemShut
  {NoStop}%
\bibitem [{\citenamefont {Xu}\ \emph {et~al.}(2016)\citenamefont {Xu},
  \citenamefont {Song}, \citenamefont {Liu}, \citenamefont {Xue}, \citenamefont
  {Su}, \citenamefont {Deng}, \citenamefont {Tian}, \citenamefont {Zheng},
  \citenamefont {Han}, \citenamefont {Zhong}, \citenamefont {Wang},
  \citenamefont {Liu},\ and\ \citenamefont {Zhao}}]{Xu2016}%
  \BibitemOpen
  \bibfield  {author} {\bibinfo {author} {\bibfnamefont {H.~K.}\ \bibnamefont
  {Xu}}, \bibinfo {author} {\bibfnamefont {C.}~\bibnamefont {Song}}, \bibinfo
  {author} {\bibfnamefont {W.~Y.}\ \bibnamefont {Liu}}, \bibinfo {author}
  {\bibfnamefont {G.~M.}\ \bibnamefont {Xue}}, \bibinfo {author} {\bibfnamefont
  {F.~F.}\ \bibnamefont {Su}}, \bibinfo {author} {\bibfnamefont
  {H.}~\bibnamefont {Deng}}, \bibinfo {author} {\bibfnamefont {Y.}~\bibnamefont
  {Tian}}, \bibinfo {author} {\bibfnamefont {D.~N.}\ \bibnamefont {Zheng}},
  \bibinfo {author} {\bibfnamefont {S.}~\bibnamefont {Han}}, \bibinfo {author}
  {\bibfnamefont {Y.~P.}\ \bibnamefont {Zhong}}, \bibinfo {author}
  {\bibfnamefont {H.}~\bibnamefont {Wang}}, \bibinfo {author} {\bibfnamefont
  {Y.-x.}\ \bibnamefont {Liu}}, \ and\ \bibinfo {author} {\bibfnamefont
  {S.~P.}\ \bibnamefont {Zhao}},\ }\href {https://doi.org/10.1038/ncomms11018}
  {\bibfield  {journal} {\bibinfo  {journal} {Nat. Commun.}\ }\textbf {\bibinfo
  {volume} {7}},\ \bibinfo {pages} {11018} (\bibinfo {year}
  {2016})}\BibitemShut {NoStop}%
\bibitem [{\citenamefont {Daems}\ and\ \citenamefont
  {Gu\'erin}(2007)}]{Daems2007}%
  \BibitemOpen
  \bibfield  {author} {\bibinfo {author} {\bibfnamefont {D.}~\bibnamefont
  {Daems}}\ and\ \bibinfo {author} {\bibfnamefont {S.}~\bibnamefont
  {Gu\'erin}},\ }\href {\doibase 10.1103/PhysRevLett.99.170503} {\bibfield
  {journal} {\bibinfo  {journal} {Phys. Rev. Lett.}\ }\textbf {\bibinfo
  {volume} {99}},\ \bibinfo {pages} {170503} (\bibinfo {year}
  {2007})}\BibitemShut {NoStop}%
\bibitem [{\citenamefont {Daems}\ and\ \citenamefont
  {Gu\'erin}(2008)}]{Daems2008}%
  \BibitemOpen
  \bibfield  {author} {\bibinfo {author} {\bibfnamefont {D.}~\bibnamefont
  {Daems}}\ and\ \bibinfo {author} {\bibfnamefont {S.}~\bibnamefont
  {Gu\'erin}},\ }\href {\doibase 10.1103/PhysRevA.78.022330} {\bibfield
  {journal} {\bibinfo  {journal} {Phys. Rev. A}\ }\textbf {\bibinfo {volume}
  {78}},\ \bibinfo {pages} {022330} (\bibinfo {year} {2008})}\BibitemShut
  {NoStop}%
\bibitem [{\citenamefont {Klein}\ \emph {et~al.}(2007)\citenamefont {Klein},
  \citenamefont {Beil},\ and\ \citenamefont {Halfmann}}]{Klein2007}%
  \BibitemOpen
  \bibfield  {author} {\bibinfo {author} {\bibfnamefont {J.}~\bibnamefont
  {Klein}}, \bibinfo {author} {\bibfnamefont {F.}~\bibnamefont {Beil}}, \ and\
  \bibinfo {author} {\bibfnamefont {T.}~\bibnamefont {Halfmann}},\ }\href
  {\doibase 10.1103/PhysRevLett.99.113003} {\bibfield  {journal} {\bibinfo
  {journal} {Phys. Rev. Lett.}\ }\textbf {\bibinfo {volume} {99}},\ \bibinfo
  {pages} {113003} (\bibinfo {year} {2007})}\BibitemShut {NoStop}%
\bibitem [{\citenamefont {Alexander}\ \emph {et~al.}(2008)\citenamefont
  {Alexander}, \citenamefont {Lauro}, \citenamefont {Louchet}, \citenamefont
  {Chaneli\`ere},\ and\ \citenamefont {Le~Gou\"et}}]{Alexander2008}%
  \BibitemOpen
  \bibfield  {author} {\bibinfo {author} {\bibfnamefont {A.~L.}\ \bibnamefont
  {Alexander}}, \bibinfo {author} {\bibfnamefont {R.}~\bibnamefont {Lauro}},
  \bibinfo {author} {\bibfnamefont {A.}~\bibnamefont {Louchet}}, \bibinfo
  {author} {\bibfnamefont {T.}~\bibnamefont {Chaneli\`ere}}, \ and\ \bibinfo
  {author} {\bibfnamefont {J.~L.}\ \bibnamefont {Le~Gou\"et}},\ }\href
  {\doibase 10.1103/PhysRevB.78.144407} {\bibfield  {journal} {\bibinfo
  {journal} {Phys. Rev. B}\ }\textbf {\bibinfo {volume} {78}},\ \bibinfo
  {pages} {144407} (\bibinfo {year} {2008})}\BibitemShut {NoStop}%
\bibitem [{\citenamefont {Vasilev}\ \emph {et~al.}(2009)\citenamefont
  {Vasilev}, \citenamefont {Kuhn},\ and\ \citenamefont
  {Vitanov}}]{Vasilev2009}%
  \BibitemOpen
  \bibfield  {author} {\bibinfo {author} {\bibfnamefont {G.~S.}\ \bibnamefont
  {Vasilev}}, \bibinfo {author} {\bibfnamefont {A.}~\bibnamefont {Kuhn}}, \
  and\ \bibinfo {author} {\bibfnamefont {N.~V.}\ \bibnamefont {Vitanov}},\
  }\href {\doibase 10.1103/PhysRevA.80.013417} {\bibfield  {journal} {\bibinfo
  {journal} {Phys. Rev. A}\ }\textbf {\bibinfo {volume} {80}},\ \bibinfo
  {pages} {013417} (\bibinfo {year} {2009})}\BibitemShut {NoStop}%
\bibitem [{\citenamefont {Dridi}\ \emph {et~al.}(2009)\citenamefont {Dridi},
  \citenamefont {Gu\'erin}, \citenamefont {Hakobyan}, \citenamefont {Jauslin},\
  and\ \citenamefont {Eleuch}}]{Dridi2009}%
  \BibitemOpen
  \bibfield  {author} {\bibinfo {author} {\bibfnamefont {G.}~\bibnamefont
  {Dridi}}, \bibinfo {author} {\bibfnamefont {S.}~\bibnamefont {Gu\'erin}},
  \bibinfo {author} {\bibfnamefont {V.}~\bibnamefont {Hakobyan}}, \bibinfo
  {author} {\bibfnamefont {H.~R.}\ \bibnamefont {Jauslin}}, \ and\ \bibinfo
  {author} {\bibfnamefont {H.}~\bibnamefont {Eleuch}},\ }\href {\doibase
  10.1103/PhysRevA.80.043408} {\bibfield  {journal} {\bibinfo  {journal} {Phys.
  Rev. A}\ }\textbf {\bibinfo {volume} {80}},\ \bibinfo {pages} {043408}
  (\bibinfo {year} {2009})}\BibitemShut {NoStop}%
\bibitem [{\citenamefont {Torosov}\ and\ \citenamefont
  {Vitanov}(2013)}]{Torosov2013}%
  \BibitemOpen
  \bibfield  {author} {\bibinfo {author} {\bibfnamefont {B.~T.}\ \bibnamefont
  {Torosov}}\ and\ \bibinfo {author} {\bibfnamefont {N.~V.}\ \bibnamefont
  {Vitanov}},\ }\href {\doibase 10.1103/PhysRevA.87.043418} {\bibfield
  {journal} {\bibinfo  {journal} {Phys. Rev. A}\ }\textbf {\bibinfo {volume}
  {87}},\ \bibinfo {pages} {043418} (\bibinfo {year} {2013})}\BibitemShut
  {NoStop}%
\bibitem [{\citenamefont {Gu\'ery-Odelin}\ \emph {et~al.}(2019)\citenamefont
  {Gu\'ery-Odelin}, \citenamefont {Ruschhaupt}, \citenamefont {Kiely},
  \citenamefont {Torrontegui}, \citenamefont {Mart\'{\i}nez-Garaot},\ and\
  \citenamefont {Muga}}]{Guery-Odelin2019}%
  \BibitemOpen
  \bibfield  {author} {\bibinfo {author} {\bibfnamefont {D.}~\bibnamefont
  {Gu\'ery-Odelin}}, \bibinfo {author} {\bibfnamefont {A.}~\bibnamefont
  {Ruschhaupt}}, \bibinfo {author} {\bibfnamefont {A.}~\bibnamefont {Kiely}},
  \bibinfo {author} {\bibfnamefont {E.}~\bibnamefont {Torrontegui}}, \bibinfo
  {author} {\bibfnamefont {S.}~\bibnamefont {Mart\'{\i}nez-Garaot}}, \ and\
  \bibinfo {author} {\bibfnamefont {J.~G.}\ \bibnamefont {Muga}},\ }\href
  {\doibase 10.1103/RevModPhys.91.045001} {\bibfield  {journal} {\bibinfo
  {journal} {Rev. Mod. Phys.}\ }\textbf {\bibinfo {volume} {91}},\ \bibinfo
  {pages} {045001} (\bibinfo {year} {2019})}\BibitemShut {NoStop}%
\bibitem [{\citenamefont {Unanyan}\ \emph {et~al.}(1997)\citenamefont
  {Unanyan}, \citenamefont {Yatsenko}, \citenamefont {Bergmann},\ and\
  \citenamefont {Shore}}]{Unanyan1997}%
  \BibitemOpen
  \bibfield  {author} {\bibinfo {author} {\bibfnamefont {R.}~\bibnamefont
  {Unanyan}}, \bibinfo {author} {\bibfnamefont {L.}~\bibnamefont {Yatsenko}},
  \bibinfo {author} {\bibfnamefont {K.}~\bibnamefont {Bergmann}}, \ and\
  \bibinfo {author} {\bibfnamefont {B.}~\bibnamefont {Shore}},\ }\href
  {\doibase https://doi.org/10.1016/S0030-4018(97)00099-0} {\bibfield
  {journal} {\bibinfo  {journal} {Opt. Commun.}\ }\textbf {\bibinfo {volume}
  {139}},\ \bibinfo {pages} {48 } (\bibinfo {year} {1997})}\BibitemShut
  {NoStop}%
\bibitem [{\citenamefont {Demirplak}\ and\ \citenamefont
  {Rice}(2003)}]{Demirplak2003}%
  \BibitemOpen
  \bibfield  {author} {\bibinfo {author} {\bibfnamefont {M.}~\bibnamefont
  {Demirplak}}\ and\ \bibinfo {author} {\bibfnamefont {S.~A.}\ \bibnamefont
  {Rice}},\ }\href {\doibase 10.1021/jp030708a} {\bibfield  {journal} {\bibinfo
   {journal} {J. Phys. Chem. A}\ }\textbf {\bibinfo {volume} {107}},\ \bibinfo
  {pages} {9937} (\bibinfo {year} {2003})}\BibitemShut {NoStop}%
\bibitem [{\citenamefont {Demirplak}\ and\ \citenamefont
  {Rice}(2008)}]{Demirplak2008}%
  \BibitemOpen
  \bibfield  {author} {\bibinfo {author} {\bibfnamefont {M.}~\bibnamefont
  {Demirplak}}\ and\ \bibinfo {author} {\bibfnamefont {S.}~\bibnamefont
  {Rice}},\ }\href {\doibase 10.1063/1.2992152} {\bibfield  {journal} {\bibinfo
   {journal} {J. Chem. Phys.}\ }\textbf {\bibinfo {volume} {129}},\ \bibinfo
  {pages} {154111} (\bibinfo {year} {2008})}\BibitemShut {NoStop}%
\bibitem [{\citenamefont {Berry}(2009)}]{Berry2009}%
  \BibitemOpen
  \bibfield  {author} {\bibinfo {author} {\bibfnamefont {M.~V.}\ \bibnamefont
  {Berry}},\ }\href {\doibase 10.1088/1751-8113/42/36/365303} {\bibfield
  {journal} {\bibinfo  {journal} {J. Phys. A: Math. Theor.}\ }\textbf {\bibinfo
  {volume} {42}},\ \bibinfo {pages} {365303} (\bibinfo {year}
  {2009})}\BibitemShut {NoStop}%
\bibitem [{\citenamefont {Chen}\ \emph {et~al.}(2010)\citenamefont {Chen},
  \citenamefont {Lizuain}, \citenamefont {Ruschhaupt}, \citenamefont
  {Gu\'ery-Odelin},\ and\ \citenamefont {Muga}}]{Chen2010}%
  \BibitemOpen
  \bibfield  {author} {\bibinfo {author} {\bibfnamefont {X.}~\bibnamefont
  {Chen}}, \bibinfo {author} {\bibfnamefont {I.}~\bibnamefont {Lizuain}},
  \bibinfo {author} {\bibfnamefont {A.}~\bibnamefont {Ruschhaupt}}, \bibinfo
  {author} {\bibfnamefont {D.}~\bibnamefont {Gu\'ery-Odelin}}, \ and\ \bibinfo
  {author} {\bibfnamefont {J.~G.}\ \bibnamefont {Muga}},\ }\href {\doibase
  10.1103/PhysRevLett.105.123003} {\bibfield  {journal} {\bibinfo  {journal}
  {Phys. Rev. Lett.}\ }\textbf {\bibinfo {volume} {105}},\ \bibinfo {pages}
  {123003} (\bibinfo {year} {2010})}\BibitemShut {NoStop}%
\bibitem [{\citenamefont {Du}\ \emph {et~al.}(2016)\citenamefont {Du},
  \citenamefont {Liang}, \citenamefont {Li}, \citenamefont {Yue}, \citenamefont
  {Lv}, \citenamefont {Huang}, \citenamefont {Chen}, \citenamefont {Yan},\ and\
  \citenamefont {Zhu}}]{Du2016}%
  \BibitemOpen
  \bibfield  {author} {\bibinfo {author} {\bibfnamefont {Y.-X.}\ \bibnamefont
  {Du}}, \bibinfo {author} {\bibfnamefont {Z.-T.}\ \bibnamefont {Liang}},
  \bibinfo {author} {\bibfnamefont {Y.-C.}\ \bibnamefont {Li}}, \bibinfo
  {author} {\bibfnamefont {X.-X.}\ \bibnamefont {Yue}}, \bibinfo {author}
  {\bibfnamefont {Q.-X.}\ \bibnamefont {Lv}}, \bibinfo {author} {\bibfnamefont
  {W.}~\bibnamefont {Huang}}, \bibinfo {author} {\bibfnamefont
  {X.}~\bibnamefont {Chen}}, \bibinfo {author} {\bibfnamefont {H.}~\bibnamefont
  {Yan}}, \ and\ \bibinfo {author} {\bibfnamefont {S.-L.}\ \bibnamefont
  {Zhu}},\ }\href {https://doi.org/10.1038/ncomms12479} {\bibfield  {journal}
  {\bibinfo  {journal} {Nat. Commun.}\ }\textbf {\bibinfo {volume} {7}},\
  \bibinfo {pages} {12479} (\bibinfo {year} {2016})}\BibitemShut {NoStop}%
\bibitem [{\citenamefont {An}\ \emph {et~al.}(2016)\citenamefont {An},
  \citenamefont {Lv}, \citenamefont {del Campo},\ and\ \citenamefont
  {Kim}}]{An2016}%
  \BibitemOpen
  \bibfield  {author} {\bibinfo {author} {\bibfnamefont {S.}~\bibnamefont
  {An}}, \bibinfo {author} {\bibfnamefont {D.}~\bibnamefont {Lv}}, \bibinfo
  {author} {\bibfnamefont {A.}~\bibnamefont {del Campo}}, \ and\ \bibinfo
  {author} {\bibfnamefont {K.}~\bibnamefont {Kim}},\ }\href
  {https://doi.org/10.1038/ncomms12999} {\bibfield  {journal} {\bibinfo
  {journal} {Nat. Commun.}\ }\textbf {\bibinfo {volume} {7}},\ \bibinfo {pages}
  {12999} (\bibinfo {year} {2016})}\BibitemShut {NoStop}%
\bibitem [{\citenamefont {Vepsalainen}\ \emph {et~al.}(2019)\citenamefont
  {Vepsalainen}, \citenamefont {Danilin},\ and\ \citenamefont
  {Paraoanu}}]{Vepsalainen2019}%
  \BibitemOpen
  \bibfield  {author} {\bibinfo {author} {\bibfnamefont {A.}~\bibnamefont
  {Vepsalainen}}, \bibinfo {author} {\bibfnamefont {S.}~\bibnamefont
  {Danilin}}, \ and\ \bibinfo {author} {\bibfnamefont {G.~S.}\ \bibnamefont
  {Paraoanu}},\ }\href
  {http://advances.sciencemag.org/content/5/2/eaau5999.abstract} {\bibfield
  {journal} {\bibinfo  {journal} {Sci. Adv.}\ }\textbf {\bibinfo {volume}
  {5}},\ \bibinfo {pages} {eaau5999} (\bibinfo {year} {2019})}\BibitemShut
  {NoStop}%
\bibitem [{\citenamefont {Yang}\ \emph {et~al.}(2019)\citenamefont {Yang},
  \citenamefont {Tan}, \citenamefont {Dong}, \citenamefont {Yang},
  \citenamefont {Song}, \citenamefont {Han}, \citenamefont {Chu}, \citenamefont
  {Li}, \citenamefont {Lan}, \citenamefont {Yu},\ and\ \citenamefont
  {Yu}}]{Yang2019}%
  \BibitemOpen
  \bibfield  {author} {\bibinfo {author} {\bibfnamefont {Z.}~\bibnamefont
  {Yang}}, \bibinfo {author} {\bibfnamefont {X.}~\bibnamefont {Tan}}, \bibinfo
  {author} {\bibfnamefont {Y.}~\bibnamefont {Dong}}, \bibinfo {author}
  {\bibfnamefont {X.}~\bibnamefont {Yang}}, \bibinfo {author} {\bibfnamefont
  {S.}~\bibnamefont {Song}}, \bibinfo {author} {\bibfnamefont {Z.}~\bibnamefont
  {Han}}, \bibinfo {author} {\bibfnamefont {J.}~\bibnamefont {Chu}}, \bibinfo
  {author} {\bibfnamefont {Z.}~\bibnamefont {Li}}, \bibinfo {author}
  {\bibfnamefont {D.}~\bibnamefont {Lan}}, \bibinfo {author} {\bibfnamefont
  {H.}~\bibnamefont {Yu}}, \ and\ \bibinfo {author} {\bibfnamefont
  {Y.}~\bibnamefont {Yu}},\ }\href {\doibase 10.1063/1.5111060} {\bibfield
  {journal} {\bibinfo  {journal} {Appl. Phys. Lett.}\ }\textbf {\bibinfo
  {volume} {115}},\ \bibinfo {pages} {072603} (\bibinfo {year} {2019})},\
  \Eprint {http://arxiv.org/abs/https://doi.org/10.1063/1.5111060}
  {https://doi.org/10.1063/1.5111060} \BibitemShut {NoStop}%
\bibitem [{\citenamefont {Zhou}\ \emph {et~al.}(2016)\citenamefont {Zhou},
  \citenamefont {Baksic}, \citenamefont {Ribeiro}, \citenamefont {Yale},
  \citenamefont {Heremans}, \citenamefont {Jerger}, \citenamefont {Auer},
  \citenamefont {Burkard}, \citenamefont {Clerk},\ and\ \citenamefont
  {Awschalom}}]{Zhou2016}%
  \BibitemOpen
  \bibfield  {author} {\bibinfo {author} {\bibfnamefont {B.~B.}\ \bibnamefont
  {Zhou}}, \bibinfo {author} {\bibfnamefont {A.}~\bibnamefont {Baksic}},
  \bibinfo {author} {\bibfnamefont {H.}~\bibnamefont {Ribeiro}}, \bibinfo
  {author} {\bibfnamefont {C.~G.}\ \bibnamefont {Yale}}, \bibinfo {author}
  {\bibfnamefont {F.~J.}\ \bibnamefont {Heremans}}, \bibinfo {author}
  {\bibfnamefont {P.~C.}\ \bibnamefont {Jerger}}, \bibinfo {author}
  {\bibfnamefont {A.}~\bibnamefont {Auer}}, \bibinfo {author} {\bibfnamefont
  {G.}~\bibnamefont {Burkard}}, \bibinfo {author} {\bibfnamefont {A.~A.}\
  \bibnamefont {Clerk}}, \ and\ \bibinfo {author} {\bibfnamefont {D.~D.}\
  \bibnamefont {Awschalom}},\ }\href {https://doi.org/10.1038/nphys3967}
  {\bibfield  {journal} {\bibinfo  {journal} {Nat. Phys.}\ }\textbf {\bibinfo
  {volume} {13}},\ \bibinfo {pages} {330} (\bibinfo {year} {2016})}\BibitemShut
  {NoStop}%
\bibitem [{\citenamefont {Baksic}\ \emph {et~al.}(2016)\citenamefont {Baksic},
  \citenamefont {Ribeiro},\ and\ \citenamefont {Clerk}}]{Baksic2016}%
  \BibitemOpen
  \bibfield  {author} {\bibinfo {author} {\bibfnamefont {A.}~\bibnamefont
  {Baksic}}, \bibinfo {author} {\bibfnamefont {H.}~\bibnamefont {Ribeiro}}, \
  and\ \bibinfo {author} {\bibfnamefont {A.~A.}\ \bibnamefont {Clerk}},\ }\href
  {\doibase 10.1103/PhysRevLett.116.230503} {\bibfield  {journal} {\bibinfo
  {journal} {Phys. Rev. Lett.}\ }\textbf {\bibinfo {volume} {116}},\ \bibinfo
  {pages} {230503} (\bibinfo {year} {2016})}\BibitemShut {NoStop}%
\bibitem [{\citenamefont {Chen}\ and\ \citenamefont {Muga}(2012)}]{Chen2012}%
  \BibitemOpen
  \bibfield  {author} {\bibinfo {author} {\bibfnamefont {X.}~\bibnamefont
  {Chen}}\ and\ \bibinfo {author} {\bibfnamefont {J.~G.}\ \bibnamefont
  {Muga}},\ }\href {\doibase 10.1103/PhysRevA.86.033405} {\bibfield  {journal}
  {\bibinfo  {journal} {Phys. Rev. A}\ }\textbf {\bibinfo {volume} {86}},\
  \bibinfo {pages} {033405} (\bibinfo {year} {2012})}\BibitemShut {NoStop}%
\bibitem [{\citenamefont {Laforgue}\ \emph {et~al.}(2019)\citenamefont
  {Laforgue}, \citenamefont {Chen},\ and\ \citenamefont
  {Gu\'erin}}]{Laforgue2019}%
  \BibitemOpen
  \bibfield  {author} {\bibinfo {author} {\bibfnamefont {X.}~\bibnamefont
  {Laforgue}}, \bibinfo {author} {\bibfnamefont {X.}~\bibnamefont {Chen}}, \
  and\ \bibinfo {author} {\bibfnamefont {S.}~\bibnamefont {Gu\'erin}},\ }\href
  {\doibase 10.1103/PhysRevA.100.023415} {\bibfield  {journal} {\bibinfo
  {journal} {Phys. Rev. A}\ }\textbf {\bibinfo {volume} {100}},\ \bibinfo
  {pages} {023415} (\bibinfo {year} {2019})}\BibitemShut {NoStop}%
\bibitem [{\citenamefont {Xu}\ \emph {et~al.}(2019)\citenamefont {Xu},
  \citenamefont {Liu}, \citenamefont {Chen},\ and\ \citenamefont
  {Gu{\'{e}}rin}}]{Xu2019}%
  \BibitemOpen
  \bibfield  {author} {\bibinfo {author} {\bibfnamefont {T.-N.}\ \bibnamefont
  {Xu}}, \bibinfo {author} {\bibfnamefont {K.}~\bibnamefont {Liu}}, \bibinfo
  {author} {\bibfnamefont {X.}~\bibnamefont {Chen}}, \ and\ \bibinfo {author}
  {\bibfnamefont {S.}~\bibnamefont {Gu{\'{e}}rin}},\ }\href {\doibase
  10.1088/1361-6455/ab49a9} {\bibfield  {journal} {\bibinfo  {journal} {J.
  Phys. B: At. Mol. Opt. Phys.}\ }\textbf {\bibinfo {volume} {52}},\ \bibinfo
  {pages} {235501} (\bibinfo {year} {2019})}\BibitemShut {NoStop}%
\bibitem [{\citenamefont {Davis}\ and\ \citenamefont
  {Pechukas}(1976)}]{Davis1976}%
  \BibitemOpen
  \bibfield  {author} {\bibinfo {author} {\bibfnamefont {J.~P.}\ \bibnamefont
  {Davis}}\ and\ \bibinfo {author} {\bibfnamefont {P.}~\bibnamefont
  {Pechukas}},\ }\href {\doibase 10.1063/1.432648} {\bibfield  {journal}
  {\bibinfo  {journal} {J. Chem. Phys.}\ }\textbf {\bibinfo {volume} {64}},\
  \bibinfo {pages} {3129} (\bibinfo {year} {1976})}\BibitemShut {NoStop}%
\bibitem [{\citenamefont {Motzoi}\ \emph {et~al.}(2009)\citenamefont {Motzoi},
  \citenamefont {Gambetta}, \citenamefont {Rebentrost},\ and\ \citenamefont
  {Wilhelm}}]{Motzoi2009}%
  \BibitemOpen
  \bibfield  {author} {\bibinfo {author} {\bibfnamefont {F.}~\bibnamefont
  {Motzoi}}, \bibinfo {author} {\bibfnamefont {J.~M.}\ \bibnamefont
  {Gambetta}}, \bibinfo {author} {\bibfnamefont {P.}~\bibnamefont
  {Rebentrost}}, \ and\ \bibinfo {author} {\bibfnamefont {F.~K.}\ \bibnamefont
  {Wilhelm}},\ }\href {\doibase 10.1103/PhysRevLett.103.110501} {\bibfield
  {journal} {\bibinfo  {journal} {Phys. Rev. Lett.}\ }\textbf {\bibinfo
  {volume} {103}},\ \bibinfo {pages} {110501} (\bibinfo {year}
  {2009})}\BibitemShut {NoStop}%
\bibitem [{\citenamefont {Gambetta}\ \emph {et~al.}(2011)\citenamefont
  {Gambetta}, \citenamefont {Motzoi}, \citenamefont {Merkel},\ and\
  \citenamefont {Wilhelm}}]{Gambetta2011}%
  \BibitemOpen
  \bibfield  {author} {\bibinfo {author} {\bibfnamefont {J.~M.}\ \bibnamefont
  {Gambetta}}, \bibinfo {author} {\bibfnamefont {F.}~\bibnamefont {Motzoi}},
  \bibinfo {author} {\bibfnamefont {S.~T.}\ \bibnamefont {Merkel}}, \ and\
  \bibinfo {author} {\bibfnamefont {F.~K.}\ \bibnamefont {Wilhelm}},\ }\href
  {\doibase 10.1103/PhysRevA.83.012308} {\bibfield  {journal} {\bibinfo
  {journal} {Phys. Rev. A}\ }\textbf {\bibinfo {volume} {83}},\ \bibinfo
  {pages} {012308} (\bibinfo {year} {2011})}\BibitemShut {NoStop}%
\bibitem [{\citenamefont {Chen}\ \emph {et~al.}(2016)\citenamefont {Chen},
  \citenamefont {Kelly}, \citenamefont {Quintana}, \citenamefont {Barends},
  \citenamefont {Campbell}, \citenamefont {Chen}, \citenamefont {Chiaro},
  \citenamefont {Dunsworth}, \citenamefont {Fowler}, \citenamefont {Lucero},
  \citenamefont {Jeffrey}, \citenamefont {Megrant}, \citenamefont {Mutus},
  \citenamefont {Neeley}, \citenamefont {Neill}, \citenamefont {O'Malley},
  \citenamefont {Roushan}, \citenamefont {Sank}, \citenamefont {Vainsencher},
  \citenamefont {Wenner}, \citenamefont {White}, \citenamefont {Korotkov},\
  and\ \citenamefont {Martinis}}]{Chen2016}%
  \BibitemOpen
  \bibfield  {author} {\bibinfo {author} {\bibfnamefont {Z.}~\bibnamefont
  {Chen}}, \bibinfo {author} {\bibfnamefont {J.}~\bibnamefont {Kelly}},
  \bibinfo {author} {\bibfnamefont {C.}~\bibnamefont {Quintana}}, \bibinfo
  {author} {\bibfnamefont {R.}~\bibnamefont {Barends}}, \bibinfo {author}
  {\bibfnamefont {B.}~\bibnamefont {Campbell}}, \bibinfo {author}
  {\bibfnamefont {Y.}~\bibnamefont {Chen}}, \bibinfo {author} {\bibfnamefont
  {B.}~\bibnamefont {Chiaro}}, \bibinfo {author} {\bibfnamefont
  {A.}~\bibnamefont {Dunsworth}}, \bibinfo {author} {\bibfnamefont {A.~G.}\
  \bibnamefont {Fowler}}, \bibinfo {author} {\bibfnamefont {E.}~\bibnamefont
  {Lucero}}, \bibinfo {author} {\bibfnamefont {E.}~\bibnamefont {Jeffrey}},
  \bibinfo {author} {\bibfnamefont {A.}~\bibnamefont {Megrant}}, \bibinfo
  {author} {\bibfnamefont {J.}~\bibnamefont {Mutus}}, \bibinfo {author}
  {\bibfnamefont {M.}~\bibnamefont {Neeley}}, \bibinfo {author} {\bibfnamefont
  {C.}~\bibnamefont {Neill}}, \bibinfo {author} {\bibfnamefont {P.~J.~J.}\
  \bibnamefont {O'Malley}}, \bibinfo {author} {\bibfnamefont {P.}~\bibnamefont
  {Roushan}}, \bibinfo {author} {\bibfnamefont {D.}~\bibnamefont {Sank}},
  \bibinfo {author} {\bibfnamefont {A.}~\bibnamefont {Vainsencher}}, \bibinfo
  {author} {\bibfnamefont {J.}~\bibnamefont {Wenner}}, \bibinfo {author}
  {\bibfnamefont {T.~C.}\ \bibnamefont {White}}, \bibinfo {author}
  {\bibfnamefont {A.~N.}\ \bibnamefont {Korotkov}}, \ and\ \bibinfo {author}
  {\bibfnamefont {J.~M.}\ \bibnamefont {Martinis}},\ }\href {\doibase
  10.1103/PhysRevLett.116.020501} {\bibfield  {journal} {\bibinfo  {journal}
  {Phys. Rev. Lett.}\ }\textbf {\bibinfo {volume} {116}},\ \bibinfo {pages}
  {020501} (\bibinfo {year} {2016})}\BibitemShut {NoStop}%
\bibitem [{\citenamefont {Christandl}\ \emph {et~al.}(2004)\citenamefont
  {Christandl}, \citenamefont {Datta}, \citenamefont {Ekert},\ and\
  \citenamefont {Landahl}}]{Christandl2004}%
  \BibitemOpen
  \bibfield  {author} {\bibinfo {author} {\bibfnamefont {M.}~\bibnamefont
  {Christandl}}, \bibinfo {author} {\bibfnamefont {N.}~\bibnamefont {Datta}},
  \bibinfo {author} {\bibfnamefont {A.}~\bibnamefont {Ekert}}, \ and\ \bibinfo
  {author} {\bibfnamefont {A.~J.}\ \bibnamefont {Landahl}},\ }\href {\doibase
  10.1103/PhysRevLett.92.187902} {\bibfield  {journal} {\bibinfo  {journal}
  {Phys. Rev. Lett.}\ }\textbf {\bibinfo {volume} {92}},\ \bibinfo {pages}
  {187902} (\bibinfo {year} {2004})}\BibitemShut {NoStop}%
\bibitem [{\citenamefont {Li}\ \emph {et~al.}(2018)\citenamefont {Li},
  \citenamefont {Ma}, \citenamefont {Han}, \citenamefont {Chen}, \citenamefont
  {Xu}, \citenamefont {Cai}, \citenamefont {Wang}, \citenamefont {Song},
  \citenamefont {Xue}, \citenamefont {Yin},\ and\ \citenamefont
  {Sun}}]{Li2018a}%
  \BibitemOpen
  \bibfield  {author} {\bibinfo {author} {\bibfnamefont {X.}~\bibnamefont
  {Li}}, \bibinfo {author} {\bibfnamefont {Y.}~\bibnamefont {Ma}}, \bibinfo
  {author} {\bibfnamefont {J.}~\bibnamefont {Han}}, \bibinfo {author}
  {\bibfnamefont {T.}~\bibnamefont {Chen}}, \bibinfo {author} {\bibfnamefont
  {Y.}~\bibnamefont {Xu}}, \bibinfo {author} {\bibfnamefont {W.}~\bibnamefont
  {Cai}}, \bibinfo {author} {\bibfnamefont {H.}~\bibnamefont {Wang}}, \bibinfo
  {author} {\bibfnamefont {Y.}~\bibnamefont {Song}}, \bibinfo {author}
  {\bibfnamefont {Z.-Y.}\ \bibnamefont {Xue}}, \bibinfo {author} {\bibfnamefont
  {Z.-q.}\ \bibnamefont {Yin}}, \ and\ \bibinfo {author} {\bibfnamefont
  {L.}~\bibnamefont {Sun}},\ }\href {\doibase 10.1103/PhysRevApplied.10.054009}
  {\bibfield  {journal} {\bibinfo  {journal} {Phys. Rev. Applied}\ }\textbf
  {\bibinfo {volume} {10}},\ \bibinfo {pages} {054009} (\bibinfo {year}
  {2018})}\BibitemShut {NoStop}%
\bibitem [{\citenamefont {Liu}\ and\ \citenamefont {Yung}(2019)}]{Liu2019a}%
  \BibitemOpen
  \bibfield  {author} {\bibinfo {author} {\bibfnamefont {B.-J.}\ \bibnamefont
  {Liu}}\ and\ \bibinfo {author} {\bibfnamefont {M.-H.}\ \bibnamefont {Yung}},\
  }\href@noop {} {\bibfield  {journal} {\bibinfo  {journal} {to be published}\
  } (\bibinfo {year} {2019})}\BibitemShut {NoStop}%
\bibitem [{\citenamefont {Panda}\ \emph {et~al.}(2016)\citenamefont {Panda},
  \citenamefont {O'Leary}, \citenamefont {West}, \citenamefont {Baron},
  \citenamefont {Hess}, \citenamefont {Hoffman}, \citenamefont {Kirilov},
  \citenamefont {Overstreet}, \citenamefont {West}, \citenamefont {DeMille},
  \citenamefont {Doyle},\ and\ \citenamefont {Gabrielse}}]{Panda2016}%
  \BibitemOpen
  \bibfield  {author} {\bibinfo {author} {\bibfnamefont {C.~D.}\ \bibnamefont
  {Panda}}, \bibinfo {author} {\bibfnamefont {B.~R.}\ \bibnamefont {O'Leary}},
  \bibinfo {author} {\bibfnamefont {A.~D.}\ \bibnamefont {West}}, \bibinfo
  {author} {\bibfnamefont {J.}~\bibnamefont {Baron}}, \bibinfo {author}
  {\bibfnamefont {P.~W.}\ \bibnamefont {Hess}}, \bibinfo {author}
  {\bibfnamefont {C.}~\bibnamefont {Hoffman}}, \bibinfo {author} {\bibfnamefont
  {E.}~\bibnamefont {Kirilov}}, \bibinfo {author} {\bibfnamefont {C.~B.}\
  \bibnamefont {Overstreet}}, \bibinfo {author} {\bibfnamefont {E.~P.}\
  \bibnamefont {West}}, \bibinfo {author} {\bibfnamefont {D.}~\bibnamefont
  {DeMille}}, \bibinfo {author} {\bibfnamefont {J.~M.}\ \bibnamefont {Doyle}},
  \ and\ \bibinfo {author} {\bibfnamefont {G.}~\bibnamefont {Gabrielse}},\
  }\href {\doibase 10.1103/PhysRevA.93.052110} {\bibfield  {journal} {\bibinfo
  {journal} {Phys. Rev. A}\ }\textbf {\bibinfo {volume} {93}},\ \bibinfo
  {pages} {052110} (\bibinfo {year} {2016})}\BibitemShut {NoStop}%
\bibitem [{\citenamefont {Dorier}\ \emph {et~al.}(2017)\citenamefont {Dorier},
  \citenamefont {Gevorgyan}, \citenamefont {Ishkhanyan}, \citenamefont {Leroy},
  \citenamefont {Jauslin},\ and\ \citenamefont {Gu\'erin}}]{Dorier2017}%
  \BibitemOpen
  \bibfield  {author} {\bibinfo {author} {\bibfnamefont {V.}~\bibnamefont
  {Dorier}}, \bibinfo {author} {\bibfnamefont {M.}~\bibnamefont {Gevorgyan}},
  \bibinfo {author} {\bibfnamefont {A.}~\bibnamefont {Ishkhanyan}}, \bibinfo
  {author} {\bibfnamefont {C.}~\bibnamefont {Leroy}}, \bibinfo {author}
  {\bibfnamefont {H.~R.}\ \bibnamefont {Jauslin}}, \ and\ \bibinfo {author}
  {\bibfnamefont {S.}~\bibnamefont {Gu\'erin}},\ }\href {\doibase
  10.1103/PhysRevLett.119.243902} {\bibfield  {journal} {\bibinfo  {journal}
  {Phys. Rev. Lett.}\ }\textbf {\bibinfo {volume} {119}},\ \bibinfo {pages}
  {243902} (\bibinfo {year} {2017})}\BibitemShut {NoStop}%
\end{thebibliography}
\end{document}